\documentclass[amssymb,useAMS,prd,aps,amsmath,superscriptaddress,nofootinbib,twocolumn]{revtex4}
\usepackage{color}
\usepackage{pslatex}
\usepackage{pdfpages}
\usepackage{graphicx}
\usepackage{psfrag}
\usepackage{pdftricks}
\usepackage{multirow}
\usepackage{graphicx}
\usepackage{hyperref}

\usepackage[normalem]{ulem}

\begin{document}
\title{A possible link between the GeV   excess and the 511 keV emission line in the Galactic Centre: Erratum}

\preprint{IPPP/14/33 DCPT/14/64, SLAC-PUB-15945}

%\date{\today}

\author{C\'eline B\oe hm}
\affiliation{Institute for Particle Physics Phenomenology, Durham University, South Road, Durham, DH1 3LE, United Kingdom}
\affiliation{LAPTH, U. de Savoie, CNRS,  BP 110, 74941 Annecy-Le-Vieux, France}
\affiliation{Nordita, KTH Royal Institute of Technology and Stockholm University, Roslagstullsbacken 23, SE-106 91 Stockholm, Sweden }
\author{Paolo Gondolo}
\affiliation{Physics And Astronomy, University of Utah, 201 South Presidents Circle Room 201, Salt Lake City, UT 84112}
\affiliation{Nordita, KTH Royal Institute of Technology and Stockholm University, Roslagstullsbacken 23, SE-106 91 Stockholm, Sweden }
\author{Pierre Jean}
\affiliation{Universit\'e de Toulouse; UPS-OMP;
IRAP; Toulouse, France}
\affiliation{CNRS; IRAP; 9 Av. colonel Roche, BP
44346, F-31028 Toulouse cedex 4,
France}
\author{Thomas Lacroix}
\affiliation{UMR7095, Institut d'Astrophysique de Paris, 98 bis boulevard Arago, 75014 Paris, France}
\affiliation{Nordita, KTH Royal Institute of Technology and Stockholm University, Roslagstullsbacken 23, SE-106 91 Stockholm, Sweden }
\author{Colin Norman}
\affiliation{Johns Hopkins University, 
Department of Physics $\&$ Astronomy, Bloomberg Center for Physics and Astronomy,  Baltimore, MD 21218, USA}
\author{Joseph Silk}
\affiliation{UMR7095, Institut d'Astrophysique de Paris, 98 bis boulevard Arago, 75014 Paris, France}
\affiliation{Johns Hopkins University, 
Department of Physics $\&$ Astronomy, Bloomberg Center for Physics and Astronomy,  Baltimore, MD 21218, USA}
%{\smallskip \tt  \footnotesize \href{mailto:c.m.boehm@durham.ac.uk}{c.m.boehm@durham.ac.uk}, \smallskip}

\begin{abstract}
The morphology and characteristics of the so-called GeV gamma-ray excess detected in the Milky Way lead us to speculate about a possible common origin with the 511 keV line 
  mapped by the SPI experiment about ten years ago. In the previous version of our paper, we assumed 30 GeV dark matter particles annihilating into $b \bar{b}$ 
 and obtained both a morphology and a 511 keV flux ($\phi_{511 \rm{keV}} \simeq 10^{-3} \ \rm{ph/cm^2/s}$)  
 in agreement with SPI observation. 
However our estimates assumed a negligible number density of electrons in the bulge which lead to an artificial increase in the flux (mostly due to negligible Coulomb losses in this configuration). 
Assuming a number density greater than $n_e > 10^{-3} \  \rm{cm^{-3}}$, we now obtain a flux of 511 keV photons that is smaller than $\phi_{511 \rm{keV}} \simeq 10^{-6} \ \rm{ph/cm^2/s}$ and is essentially in agreement with the 511 keV flux that one can infer from the total number of positrons injected by dark matter annihilations 
into $b \bar{b}$. We thus conclude that -- even if 30 GeV dark matter particles were to exist-- it is impossible to establish a connexion between the two types of signals, even though they are located within the same 10 deg region in the galactic centre. 
 \end{abstract}

 \maketitle

\section{Total number of positrons at injection}

The flux of 511 keV photons which has been reported by the SPI experiment is about $\phi_{511} \simeq 10^{-3}\  \rm{ph/cm^2/s}$. This number corresponds to a luminosity of about 
$L_{511} \simeq 10^{43}\ \rm{ph/s}$. As these 511 keV photons are supposed to originate from positronium formation, the number of positrons at energies below 100 eV (where positronium 
formation becomes efficient) is about $L_{e^+} \simeq L_{511} /2 \sim 5 \ 10^{42}\ \rm  e^{+}/s$.

To determine whether such a large number of positrons can indeed originate from 30 GeV DM particles, one can first estimate the total number of positrons at injection, namely: 
$$\frac{d n_{e^+}}{d t} \vert_{inj} = \sigma v \times \frac{\rho_{\rm{DM}}^2}{m_{\rm{DM}}^2} \times {\cal{M}} \times  \eta. $$
If their sole origin is dark matter, one expects the number of positrons to be conserved after diffusion, which should therefore give a fair estimate of the maximal flux of 511 keV photons to be expected from 30 GeV dark matter particles (assuming that there is indeed a GeV excess of dark matter origin in the Galactic Centre). 

In this expression ${\cal{M}} = \int dE \ \frac{dN}{d E}  $ represents the number of positrons eventually generated by dark matter annihilations into $b \bar{b}$ and $\eta$ is a factor that accounts for the nature of the dark matter particle.  Using the spectrum provided in Ref.~\cite{Cirelli}, we find $ {\cal{M}}  \sim 10$ for a 30 GeV dark matter particle annihilating into $b \bar{b}$, leading to a number of positrons in a sphere of $10 \ \rm{kpc}$ of about:

\begin{eqnarray}
\frac{d N_{e^+}}{d t} \vert_{inj} &\simeq& 1.8 \ 10^{40} \ \left(\frac{ \sigma  v}{3 \ 10^{-26} \rm{cm^3/s}}\right) \ \times \ \ \left(\frac{m_{\rm{DM}}}{\rm{30 \ GeV}} \right)^{-2} \nonumber \\
&& \hspace{1cm} \times \ \left( \frac{r_s}{30 \ \rm{kpc}}\right)^2 \ \times \ \left(\frac{{\cal{M}}}{10} \right) \ \rm{s^{-1}}. \nonumber
 \nonumber
\end{eqnarray}
For this estimate, we have taken $\eta=1/2$ (which corresponds to a Majorana dark matter particle) and considered a NFW profile with $r_\odot = 8.25$ kpc, $\gamma = 1.2$, $\rho_\odot =0.36 \ \rm{GeV/cm^3}$, as is required to explain the GeV excess. As one can readily see this result  indicates that there are not enough positrons injected by such dark matter candidates to explain the 511 keV line.

It should be stressed that at such low energies,  
it is very likely that the energy spectrum that we use and which was obtained using the Pythia Monte Carlo code is not very reliable. However in absence of a more robust prediction by other Monte Carlo codes, this provides a good enough framework for making estimates.

\begin{figure}[h]
 \centering
 \includegraphics[width=9cm]{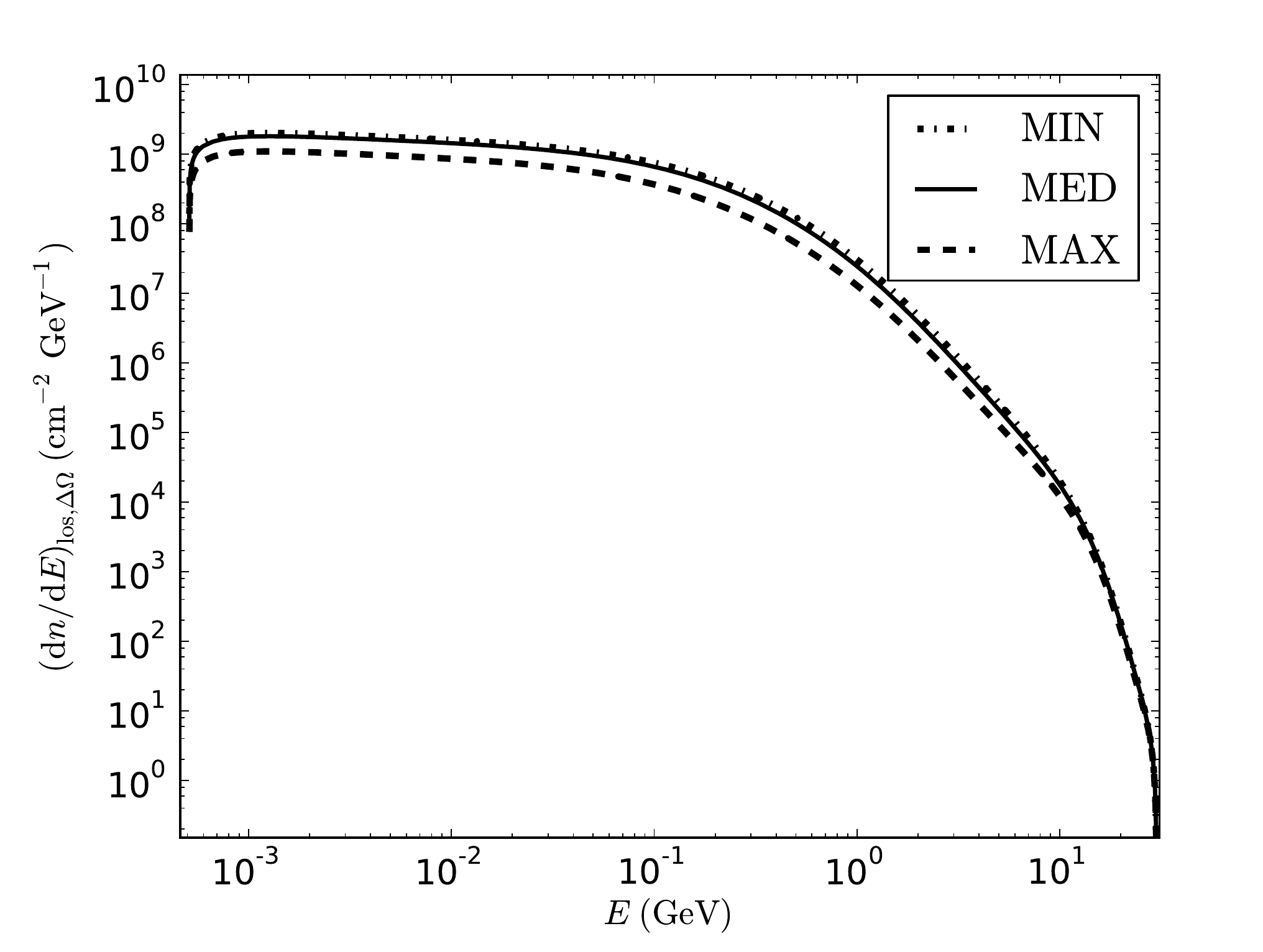}
 \caption{Diffused spectrum of $e^+$  from 30 GeV dark matter candidate annihilating into $b \ \bar{b}$ with $\sigma v = 3 \ 10^{-26} \ \rm{cm^3/s}$, using the 'MIN' , 'MED', 'MAX'    parameter sets~\cite{Delahaye:2007fr,Lacroix:2014eea}, $B= 3 \ \mu G$ and $n_H = 2 \ \rm{cm^{-3}}$. }
 \label{fig:afterdiffusion}
 \end{figure}

\section{Expected 511 keV flux}

From this estimate of the number of positrons at injection ($\frac{d N_{e^+}}{d t} \vert_{inj} \simeq 1.8 \ 10^{40} \rm{s^{-1}}$),  we deduce that even if \textbf{all} the positrons initially injected by dark matter annihilations were to lose \textbf{all} their energy, the number of 511 keV photons is 
a factor 1000 below what is required to explain the observed flux and thus leads to  $\phi_{511} \simeq 10^{-6}\  \rm{ph/cm^2/s}$. Assuming an electron number density in the bulge of 
$n_e = 10^{-3} \ \rm{cm^{-3}}$, we do actually obtain a  flux of 511 keV photons after diffusion using our code that is in agreement with the analytical estimate above. 
Typically we find the  
distribution shown in Fig.~\ref{fig:afterdiffusion}, where we note that there is no longer a sharp increase at low energy unlike Fig 2 of our version v1.

The reason for this change of behaviour at low energy is that the value $n_e \sim 10^{-3} \ \rm{cm^{-3}}$ is large enough to prevent Coulomb losses to become completely negligible,  therefore enabling the 
positrons to continue losing  energy. On the contrary, in absence of Coulomb losses ($n_e \rightarrow 0$), the positrons accumulate at very low energies, thus leading to a sharp feature in the positron distribution after diffusion at low energy. Although this rise in the positron spectrum after difussion can be easily understood, it turns out to be unphysical for the following reason: 
the number of positrons is conserved but the luminosity that results from this accumulation  exceeds the initial luminosity. 
Therefore such a sharp increase cannot exist. In other words, it is likely that in the limit $n_e \rightarrow 0$, the losses are so small that one cannot use the diffusion equation anymore 
and one needs instead to use a different technique such as the one proposed in \cite{Jean:2009zj,Alexis:2014rba}. We thus conclude that it is not possible to relate the 511 keV and the GeV excess if the origin is $b \bar{b}$ injection around 30 GeV.

\textbf{Acknowledgment}
CB would like to thank P. Richardson for useful discussions.

\bibliography{erratum_2_CBTL.bib}

\begin{thebibliography}{5}
\expandafter\ifx\csname natexlab\endcsname\relax\def\natexlab#1{#1}\fi
\expandafter\ifx\csname bibnamefont\endcsname\relax
  \def\bibnamefont#1{#1}\fi
\expandafter\ifx\csname bibfnamefont\endcsname\relax
  \def\bibfnamefont#1{#1}\fi
\expandafter\ifx\csname citenamefont\endcsname\relax
  \def\citenamefont#1{#1}\fi
\expandafter\ifx\csname url\endcsname\relax
  \def\url#1{\texttt{#1}}\fi
\expandafter\ifx\csname urlprefix\endcsname\relax\def\urlprefix{URL }\fi
\providecommand{\bibinfo}[2]{#2}
\providecommand{\eprint}[2][]{\url{#2}}

\bibitem[{\citenamefont{{Cirelli} et~al.}(2011)\citenamefont{{Cirelli},
  {Corcella}, {Hektor}, {H{\"u}tsi}, {Kadastik}, {Panci}, {Raidal}, {Sala}, and
  {Strumia}}}]{Cirelli}
\bibinfo{author}{\bibfnamefont{M.}~\bibnamefont{{Cirelli}}},
  \bibinfo{author}{\bibfnamefont{G.}~\bibnamefont{{Corcella}}},
  \bibinfo{author}{\bibfnamefont{A.}~\bibnamefont{{Hektor}}},
  \bibinfo{author}{\bibfnamefont{G.}~\bibnamefont{{H{\"u}tsi}}},
  \bibinfo{author}{\bibfnamefont{M.}~\bibnamefont{{Kadastik}}},
  \bibinfo{author}{\bibfnamefont{P.}~\bibnamefont{{Panci}}},
  \bibinfo{author}{\bibfnamefont{M.}~\bibnamefont{{Raidal}}},
  \bibinfo{author}{\bibfnamefont{F.}~\bibnamefont{{Sala}}}, \bibnamefont{and}
  \bibinfo{author}{\bibfnamefont{A.}~\bibnamefont{{Strumia}}},
  \bibinfo{journal}{JCAP} \textbf{\bibinfo{volume}{3}}, \bibinfo{pages}{51}
  (\bibinfo{year}{2011}), \bibinfo{note}{arXiv:1012.4515}.

\bibitem[{\citenamefont{Delahaye et~al.}(2008)\citenamefont{Delahaye, Lineros,
  Donato, Fornengo, and Salati}}]{Delahaye:2007fr}
\bibinfo{author}{\bibfnamefont{T.}~\bibnamefont{Delahaye}},
  \bibinfo{author}{\bibfnamefont{R.}~\bibnamefont{Lineros}},
  \bibinfo{author}{\bibfnamefont{F.}~\bibnamefont{Donato}},
  \bibinfo{author}{\bibfnamefont{N.}~\bibnamefont{Fornengo}}, \bibnamefont{and}
  \bibinfo{author}{\bibfnamefont{P.}~\bibnamefont{Salati}},
  \bibinfo{journal}{Phys.Rev.} \textbf{\bibinfo{volume}{D77}},
  \bibinfo{pages}{063527} (\bibinfo{year}{2008}), \eprint{0712.2312}.

\bibitem[{\citenamefont{Lacroix et~al.}(2014)\citenamefont{Lacroix, Boehm, and
  Silk}}]{Lacroix:2014eea}
\bibinfo{author}{\bibfnamefont{T.}~\bibnamefont{Lacroix}},
  \bibinfo{author}{\bibfnamefont{C.}~\bibnamefont{Boehm}}, \bibnamefont{and}
  \bibinfo{author}{\bibfnamefont{J.}~\bibnamefont{Silk}}
  (\bibinfo{year}{2014}), \eprint{1403.1987}.

\bibitem[{\citenamefont{Jean et~al.}(2009)\citenamefont{Jean, Gillard,
  Marcowith, and Ferriere}}]{Jean:2009zj}
\bibinfo{author}{\bibfnamefont{P.}~\bibnamefont{Jean}},
  \bibinfo{author}{\bibfnamefont{W.}~\bibnamefont{Gillard}},
  \bibinfo{author}{\bibfnamefont{A.}~\bibnamefont{Marcowith}},
  \bibnamefont{and} \bibinfo{author}{\bibfnamefont{K.}~\bibnamefont{Ferriere}}
  (\bibinfo{year}{2009}), \eprint{0909.4022}.

\bibitem[{\citenamefont{Alexis et~al.}(2014)\citenamefont{Alexis, Jean, Martin,
  and Ferriere}}]{Alexis:2014rba}
\bibinfo{author}{\bibfnamefont{A.}~\bibnamefont{Alexis}},
  \bibinfo{author}{\bibfnamefont{P.}~\bibnamefont{Jean}},
  \bibinfo{author}{\bibfnamefont{P.}~\bibnamefont{Martin}}, \bibnamefont{and}
  \bibinfo{author}{\bibfnamefont{K.}~\bibnamefont{Ferriere}}
  (\bibinfo{year}{2014}), \eprint{1402.6110}.

\end{thebibliography}

\end{document}